\documentclass[fleqn,usenatbib]{mnras}

\usepackage{newtxtext,newtxmath}

\usepackage[T1]{fontenc}

\DeclareRobustCommand{\VAN}[3]{#2}
\let\VANthebibliography\thebibliography
\def\thebibliography{\DeclareRobustCommand{\VAN}[3]{##3}\VANthebibliography}

\newcommand{\soutPC}{\bgroup\markoverwith{\textcolor{cyan}{\rule[0.5ex]{2pt}{1pt}}}\ULon}

\newcommand{\soutdif}{\bgroup\markoverwith{\textcolor{magenta}{\rule[0.5ex]{2pt}{1pt}}}\ULon}

\usepackage{graphicx}	
\usepackage{amsmath}	
\usepackage{caption}
\usepackage{xcolor}
\usepackage{subcaption}
\usepackage[normalem]{ulem} 
\usepackage{orcidlink}

\title[Cosmic wallflowers]{Cosmic wallflowers: the circumgalactic origins of isolated ultra-compact star clusters at $z>7$}

\author[F. van Donkelaar et al.] 
{Floor van Donkelaar$^{\orcidlink{0000-0002-7235-9747}}$,$^{1, 2, 3}$\thanks{E-mail: fv294@cam.ac.uk} 
Lucio Mayer$^{\orcidlink{0000-0002-7078-2074}}$,$^{1}$ 
Pedro R. Capelo$^{\orcidlink{0000-0002-1786-963X}}$,$^{1}$ 
Debora Sijacki$^{\orcidlink{0000-0002-3459-0438}}$$^{2, 3}$ and Angela Adamo$^{\orcidlink{0000-0002-8192-8091}}$$^{4}$
\\
$^{1}$Department of Astrophysics, University of Zurich, Winterthurerstrasse 190, CH-8057 Z{\"u}rich, Switzerland\\
$^{2}$Institute of Astronomy, University of Cambridge, Madingley Road, Cambridge CB3 0HA, UK\\
$^{3}$Kavli Institute for Cosmology, Cambridge (KICC), University of Cambridge, Madingley Road, Cambridge CB3 0HA, UK\\
$^{4}$Department of Astronomy, The Oskar Klein Centre, Stockholm University, AlbaNova, SE-10691 Stockholm, Sweden}

\date{Accepted XXX. Received YYY; in original form ZZZ}

\pubyear{2026}

\begin{document}
\label{firstpage}
\pagerange{\pageref{firstpage}--\pageref{lastpage}}
\maketitle

\begin{abstract}
The discovery of gravitationally lensed stellar clusters at high redshift with the James Webb Space Telescope (JWST) has revealed extremely compact, massive star-forming systems in galaxies at $z > 6$, providing a new window into early cluster formation. In this work, we investigate star cluster formation in the circumgalactic environments of gas-rich galaxies with stellar masses spanning between $\sim$$10^{8}$–$10^{11}$~M$_{\sun}$ at $z > 7$, using the \textsc{MassiveBlackPS} cosmological hydrodynamical simulation with 2~pc resolution. We identify 55 baryon-dominated clusters forming outside galactic discs but within the virial radius of the primary halo. Star formation in these systems proceeds rapidly, reaching peak stellar surface densities above $10^{5}$~M$_{\sun}$~pc$^{-2}$, closely matching the compact clusters recently discovered by JWST in the lensed Cosmic Gems Arc at $z \approx 9.6$. Such extreme densities are a key pre-requisite to trigger runaway stellar collisions, indicating that a subset of our clusters would be a likely host of intermediate-mass black holes (IMBHs). We find that massive star clusters can form efficiently in the circumgalactic medium at early times through filament fragmentation, whereby  high gas 
densities lead to rapid local collapse via a combination of thermal and gravitational instabilities. This formation pathway implies that some compact clusters formed in the quiet outskirts of forming galaxies rather than within their discs. Small variations in filament properties, including metallicity, density, and dark-matter content, influence the likelihood of a star cluster being able to form an IMBH seed. The formation of clusters in circumgalactic environments points to a potential evolutionary pathway connecting early off-disc clusters, present-day globular clusters, and the seeds of massive BHs.
\end{abstract} 

\begin{keywords}
galaxies: high-redshift -- galaxies: clusters: general -- globular clusters: general -- methods: numerical 
\end{keywords}


\section{Introduction}

Stars predominantly form in clusters, which span a wide range of ages and masses, from young open clusters to ancient globular clusters (GCs). As bound stellar systems appear to be common throughout cosmic history, constraining their formation is critical for understanding the physical conditions present during the earliest phases of galaxy assembly \citep{Brodie:2006aa}.

The advent of the James Webb Space Telescope \citep[JWST;][]{Gardner:2006aa} is revolutionizing our understanding of the early stages of galaxy formation, black holes (BHs), and stellar systems. Amongst the most surprising revelations is the discovery of remarkably dense and compact star clusters at redshifts $z \gtrsim 6$--10, many detected through gravitational lensing \citep[e.g.][]{Vanzell:2023aa, Adamo:2024aa, Mowla:2024aa, bradac:2025aa}. These clusters exhibit structural properties reminiscent of GCs. Although they are not necessarily direct progenitors of today’s GC population, their compactness and high surface densities provide key insights into clustered star formation (SF) under extreme early-Universe conditions. These observations suggest that star clusters may have been a fundamental mode of stellar assembly in the early Universe, especially as they can comprise a substantial fraction of the total stellar mass of their host galaxies \citep{adamo2025}.

At these redshifts, JWST observations show that faint galaxies in the first few hundred million years often assemble much of their stellar mass within extremely compact systems \citep[e.g.][]{Finkelstein:2024aa, Mowla:2024aa, Adamo:2024aa, Messa:2025aa}. The most compelling examples come from strongly lensed fields, where bound clusters have been identified at $z>9$, including those in the Cosmic Gems Arc. These detections confirm that gravitationally bound stellar systems formed efficiently under the physical conditions of the reionization era \citep[][]{Adamo:2024aa, Messa:2025aa}. The dense star clusters uncovered by JWST could represent the earliest stages of nuclear assembly, with mergers and feedback-regulated accretion driving the build up of galactic cores \citep[see, e.g.][]{Rantala:2024aa, Rantala:2025aa}. The coexistence of compact clusters and quenched or post-burst stellar populations observed at high redshift \citep[e.g.][]{Finkelstein:2024aa, Messa:2025aa} strengthens the view that cluster formation, feedback, and the rapid cycling of gas supply shaped the earliest phases of galaxy growth.

Observations also point to high SF efficiencies, dense interstellar media, and bursty or cyclic SF histories \citep[e.g.][]{Tacchella:2016aa, Tacchella:2020aa, Faucher:2018aa, Sun:2023aa, Sun:2023ab, Garcia:2023aa, Looser:2025aa}, reinforcing the view that clustered SF and rapid feedback cycles were a dominant mode of early stellar assembly.

In the local Universe, GCs represent the ancient end of the star cluster population and thus provide important boundary conditions on early galaxy formation. Despite this, the formation of compact stellar systems at $z \gtrsim 4$ remains poorly understood, and their origins during the first billion years are still debated \citep{Peng:2006aa, Brodie:2006aa, Forbes:2018aa}. Linking the compact clusters observed at high redshift with the GC populations seen today therefore will also require simulations capable of following cluster formation across cosmic environments.

Because the physical origin of these systems is still uncertain, a wide range of formation scenarios has been proposed. Classical models invoke formation in dark-matter (DM) haloes, mergers, or thermal instabilities \citep[e.g.][]{Peebles:1984aa, Fall:1985aa, Ashman:1922aa}, but they struggle to reconcile certain observations, such as the lack of DM in GCs today \citep{Conroy:2011aa, Ibata:2013aa}. An alternative view suggests that dense clusters like GCs formed through baryonic instabilities alone, without the need for DM. In particular, \citet{Lake:2022aa} proposes that supersonically induced gas objects (SIGOs) could seed star cluster formation. This concept aligns with the findings of \citet{vandonkelaar:2023aa}, who show that proto-GCs resembling blue GCs today can form within accreting gas filaments in the circumgalactic medium \citep[CGM; see also][]{Mandelker:2018aa}.

Building on these results, new theoretical frameworks emphasize the universality of clustered SF in high-redshift galaxies, where extreme gas surface densities, turbulence, and radiation pressure promote rapid conversion of gas into compact, bound systems \citep[e.g.][]{He:2019aa, Grudic:2021aa, Garcia:2025aa}. At the same time, observations show substantial diversity in the spatial distribution of SF at early times, with some galaxies hosting highly compact clusters while others exhibit more diffuse or irregular structures. Many of the most compact systems at $z>5$ are identified in strongly magnified JWST fields, where lensing reveals parsec-scale clustering, though unlensed compact red and dusty systems also frequently show clumpy ultraviolet (UV) emission.  The fraction of clumpy galaxies increases towards higher redshift, indicating that, while clustered SF is not universal, it becomes increasingly common in the early Universe \citep[e.g.][]{Kartaltepe:2023aa, delavega:2025}. These high clustered environments naturally produce dense clusters capable of surviving for gigayears, thereby linking systems observed at $z>9$ to present-day GCs \citep[e.g.][]{Kruijssen:2015aa, Mowla:2022aa, Adamo:2024aa}. Together, these observational and theoretical insights motivate the need for high-resolution simulations to directly test whether early compact clusters can evolve into the GC populations we observe today.

For most stellar clumps at high redshift, disc fragmentation remains a promising formation channel, particularly in environments characterized by high gas fractions and strong turbulence. The susceptibility of galactic discs to fragment depends on several ingredients, including their gas content, modality of stellar feedback, and gravitational stability \citep[e.g.][]{Dekel:2009aa, Tamburello:2015aa, Mayer:2016aa, Renaud:2021aa, vanDonkelaar:2022aa, Oklopvcic:2017aa}. Recent observations with JWST and the Atacama Large Millimeter/submillimeter Array \citep[ALMA; ][]{Wootten_Thompson_2009} suggest that rotationally supported discs may have been more prevalent at early epochs than previously thought \citep{Smit:2018aa, Rizzo:2020aa, Rizzo:2021aa, Ferreira:2022aa, Ferreira:2023aa, Roman:2023aa, Danhaive:2025aa}, while simulations indicate that fragmentation can also arise from compressive turbulence beyond linear \citet{Toomre:1964aa} instability \citep[e.g.][]{Ginzburg:2025aa}. In the \textsc{MassiveBlackPS} simulation, \citet{Mayer:2025aa} identified compact clusters with masses of $10^5$--$10^8$~M$_{\sun}$ and radii of only a few parsecs, reaching surface densities above $10^5$~M$_{\sun}$~pc$^{-2}$. Their properties closely resemble those of compact systems recently uncovered in the Cosmic Gems Arc.

Yet, there are growing indications that some compact stellar systems may be offset from main galactic discs in lensed high-redshift galaxies \citep[e.g.][]{Mowla:2022aa, Claeyssens:2023aa, Giunchi:2025aa, Whitaker:2025aa}. Such observations hint that disc fragmentation may not be the sole channel for cluster formation, and that alternative pathways, perhaps involving in-situ SF within CGM filaments, could also play a role in assembling the early cluster population \citep[e.g.][]{Mandelker:2017aa, Mandelker:2018aa, vandonkelaar:2023aa}. In addition, galaxy mergers and interactions are well known to promote vigorous cluster formation, as seen both in local systems such as the Antennae and in dedicated simulations \citep[e.g.][]{Whitmore:1999aa, Whitmore:2010aa, Renaud:2015aa, Adamo:2020aa, Lahen:2022aa}, suggesting that such processes may also have been important at high redshift. Recent observations also indicate that SF can occur within dense, clumpy galactic outflows \citep[e.g.][]{Ong:2025aa}, providing yet another potential channel for early cluster assembly.

Moreover, early cluster formation may be linked with nuclear assembly and the emergence of compact central structures. Extreme surface densities, the abundance of ``little red dots'' (LRDs), and star-by-star simulations of GC progenitors showing the formation of intermediate-mass BHs \citep[IMBHs; e.g.][]{Fujii:2024aa, Rantala:2024aa, Rantala:2025aa} support a connection between clustered SF, nuclear build-up, and BH seeding.

Motivated by these results, in this work we broaden our investigation beyond galactic disc fragmentation in the \textsc{MassiveBlackPS} simulation. We focus on the formation of dense stellar systems within the accreting gas filaments that funnel gas into the central galaxy, the ``cosmic wallflowers'', as potential  proto-GCs or BH precursors. Section~\ref{sec:method} summarizes the simulation framework and cluster identification; Section~\ref{sec:results} presents the properties of the filamentary clusters and their environments; and Section~\ref{sec:disc} discusses implications for the origins of GCs and early galaxy assembly.

\section{Methods}\label{sec:method}

\subsection{\textsc{MassiveBlackPS}}

We analyze the formation and evolution of dense, gravitationally bound star clusters in extreme high-redshift environments using the high-resolution cosmological hydrodynamical simulation \textsc{MassiveBlackPS} \citep[][]{mayer:2023aa}. The simulation was carried out with the smoothed-particle hydrodynamics (SPH) and $N$-body code \textsc{Gasoline2} \citep[][]{Wadsley:2017aa}. SF and stellar feedback were modelled following the recipes of \citet{Stinson:2006aa}, assuming a \citet{Kroupa:2001aa} initial mass function. The abundances and cooling of H and He species were calculated in non-equilibrium in the presence of a redshift-dependent radiation background \citep[][]{Haardt:2012aa}, assuming no self-shielding \citep[see][]{Pontzen:2008aa}. The cooling from the fine structure of metals was instead computed assuming photo-ionization equilibrium with the same background \citep[also assuming no self-shielding; see the discussion in][]{Capelo:2018aa} and posited to be linearly proportional to the gas metallicity \citep[][]{Shen:2010aa, Shen:2013aa}.

\begin{figure*}
    \centering
    \includegraphics[ trim={0cm 0cm 0cm 0cm}, clip, width=1\textwidth, keepaspectratio]{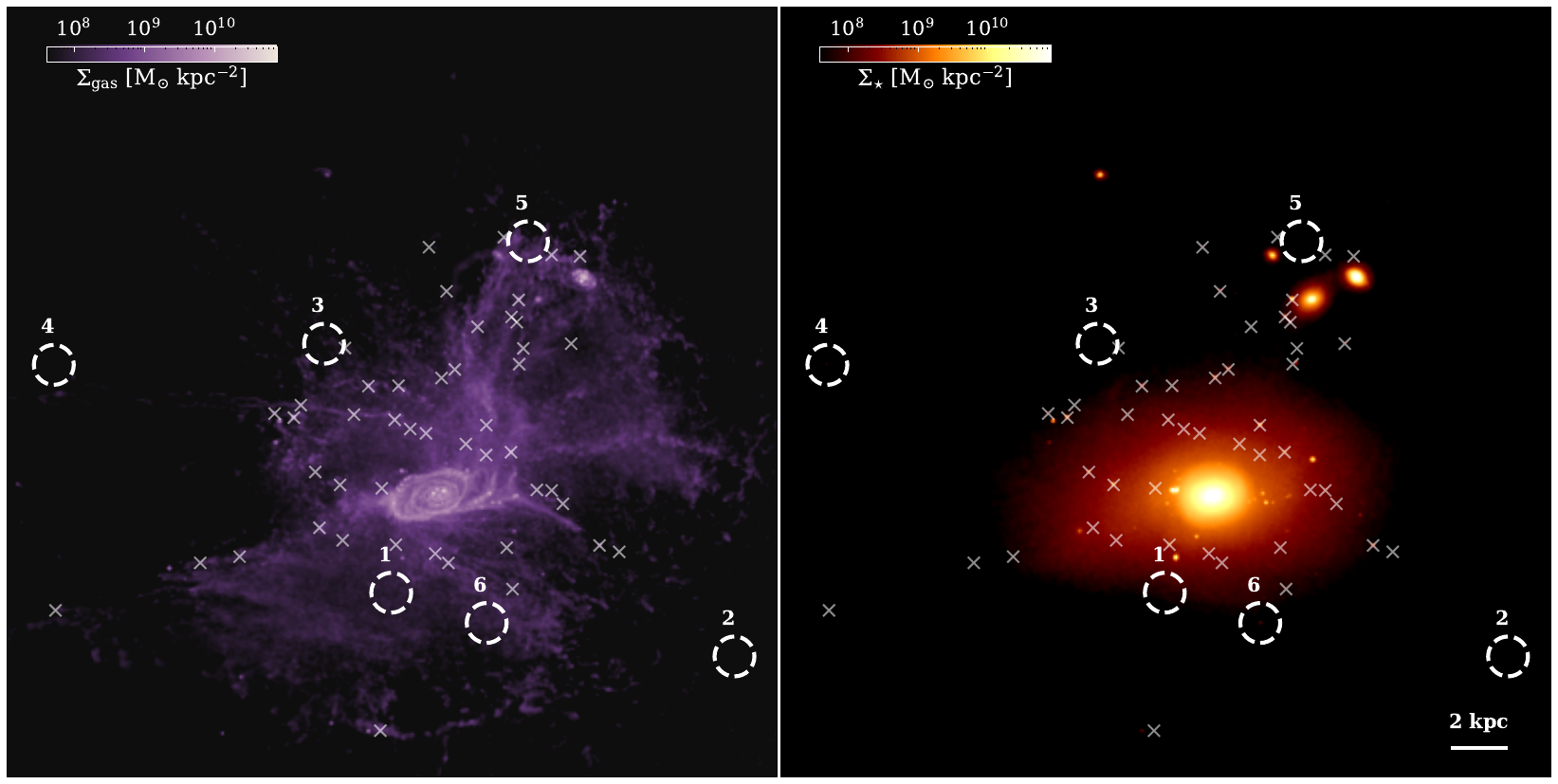}
    \caption{Gas (left-hand panel) and stellar (right-hand panel) surface density maps of the simulated system at the final snapshot, $z \sim 7.6$. The maps show the full line-of-sight projection through the simulation volume. White crosses and numbered circles mark all identified clusters that satisfy the selection criteria described in the text, with the circles highlighting six randomly selected clusters (labeled 1--6) that are zoomed-in in Figure~\ref{fig:showclust2}.}
    \label{fig:showclust1}
\end{figure*}

\textsc{MassiveBlackPS} was originally designed to explore the direct ``dark collapse'' pathway for supermassive BH (SMBH) formation. In this work, however, we shift our focus towards the emergence of isolated, or ``lonely'', compact star clusters that arise within the turbulent, merger-driven environment at $z \sim 7.6$. The simulation is based on a re-sampled sub-volume of the large-scale cosmological simulation \textsc{MassiveBlack} \citep[][]{DiMatteo_et_al_2012,Feng_et_al_2014}, which follows a $\sim$0.7~Gpc comoving volume. The target region corresponds to a massive, highly biased ($\sim$4–5$\sigma$) halo at $z \simeq 6$--7, selected to represent a typical host of a bright quasar at $z > 6$. This region undergoes repeated ``zoom-in'' refinements, ultimately reaching gas and DM particle masses of $1.9 \times 10^4$ M$_{\sun}$ and $9.4 \times 10^4$ M$_{\sun}$, respectively, with corresponding gravitational softenings of 142~pc and 241~pc. To achieve even finer resolution in the central merger remnant, \citet{mayer:2023aa} applied gas-phase particle splitting \citep[][]{Roskar_et_al_2015}, reaching a final gas mass resolution of $2.4 \times 10^3$ M$_{\sun}$ and a minimum gas gravitational softening of 2~pc in the final, particle-split isolated re-simulation. The SPH smoothing length adaptively decreases to similar or smaller scales in dense regions, enabling robust tracking of gas fragmentation and gravitational collapse \citep[][]{Bate_Burkert_1997}.

The final re-simulated volume follows only a short evolutionary window of 60~Myr, centred around the final stages of the major ($\sim$$1:1.2$) merger experienced by the primary galaxy. The isolated re-simulated region corresponds to a spherical volume of radius equal to the virial radius of the primary halo at $z \simeq 8$, with r$_{\rm vir} \approx 39$ kpc (physical). Our focus in this work is on the short evolutionary phase after the merger has ended, at $z = 7.6$, lasting about 6~Myr.

\subsection{Identifying clusters}\label{sec:structures}

\begin{figure*}
    \centering
    \includegraphics[ trim={0cm 0cm 0cm 0cm}, clip, width=0.99\textwidth, keepaspectratio]{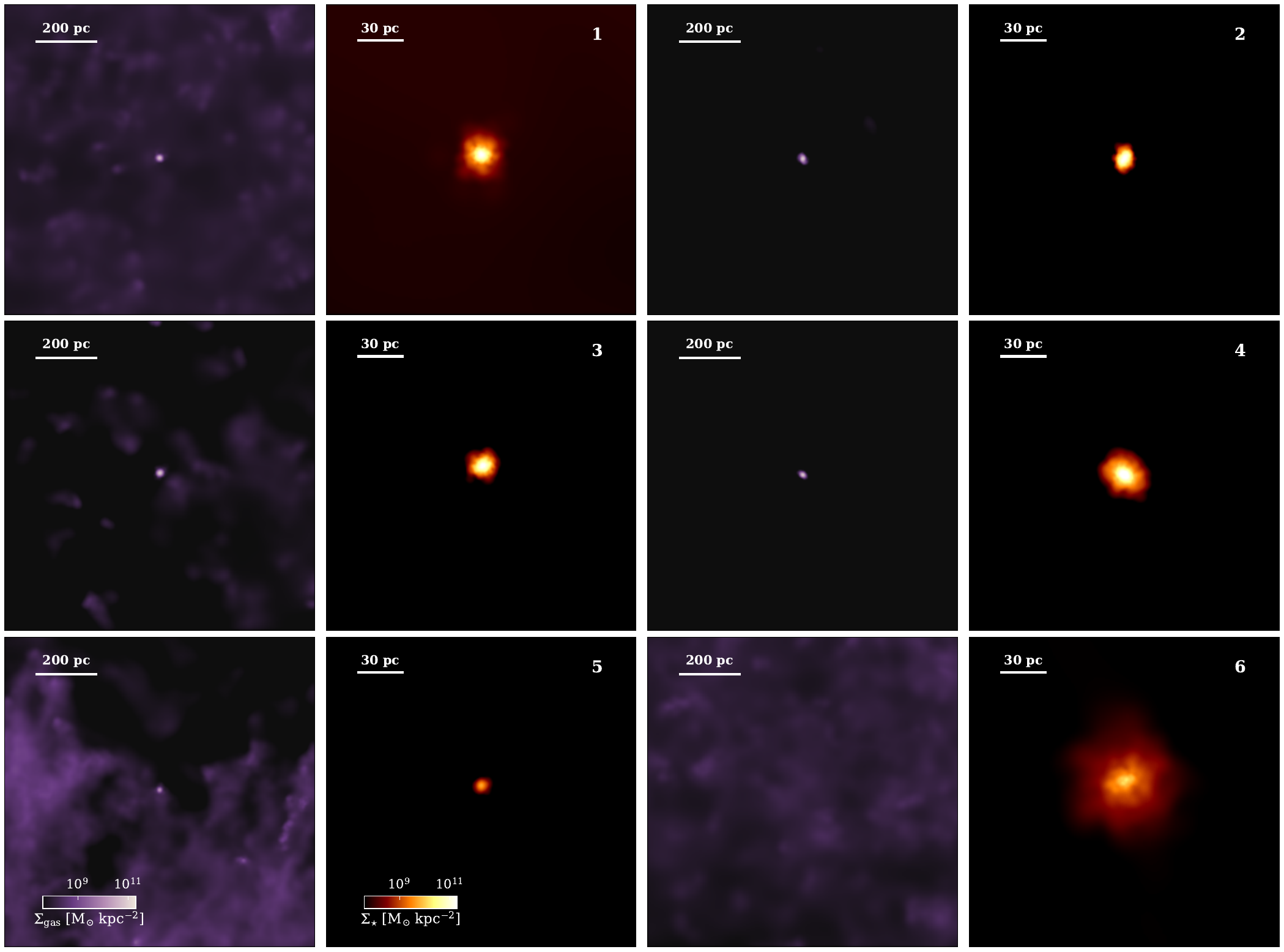}
    \caption{Zoom-in views of the six randomly selected clusters highlighted in Figure~\ref{fig:showclust1} with white, dashed, numbered circles. Each pair of panels shows the gas (left-hand side) and stellar (right-hand side) surface density maps centred on the cluster positions. In each panel, the left-hand side shows the gas surface density within a (1~kpc)$^2$ region, whereas the right-hand side presents a zoomed-in view of the stellar surface density within a (200~pc)$^2$ region.}
    \label{fig:showclust2}
\end{figure*}

We identified stellar substructures with the adaptive-mesh \textsc{AMIGA Halo Finder} \citep[\textsc{AHF};][]{Gill:2004aa, Knollmann:2009aa}. Candidate clusters are required to lie at least 2.5~kpc from the centre of the central halo (Halo~0), a distance chosen because the stellar density in the inner region is so high that it becomes difficult to unambiguously determine whether a compact object is physically associated with the central galaxy \citep[a constrain also discussed in][with many of the low-mass clusters missing in the central halo]{Mayer:2025aa}. Clusters are also required not to overlap with any other identified halo; this condition was first verified using the AHF halo catalogues, by ensuring that none of the candidates are identified as substructures of a secondary halo, and then further confirmed through visual inspection. 

To ensure that the clusters are numerically resolved, we required each system to have a bound structure with an effective radius from a \citet{king:1962aa}-model fit to the projected stellar density profile (i.e. the projected radius enclosing half of the model stellar mass) exceeding at least four times the gravitational softening of 2~pc\footnote{The lowest stellar effective radius of the identified clusters is 9.5~pc, well above the gravitational softening of 2~pc.}. Furthermore, to exclude dwarf galaxies, we applied strict upper limits on the cluster properties: the stellar half-mass radius must be smaller than 15~pc, and the total stellar mass must be less than $8 \times 10^7$~M$_{\sun}$. These limits are guided by the observed structural properties of compact clusters and dwarf galaxies in the local Universe: GC typically have half-light radii up to $\sim$10--15~pc even at the high-mass end \citep{Longeard:2019aa}, whereas dwarfs generally occupy larger sizes and higher masses \citep{McConnachie:2012aa,Willman:2005aa}. Finally, we verified that none of the selected systems are DM-dominated, all candidates exhibit negligible bound DM content, as expected given the above criteria. Specifically, we excluded any object with a bound baryon mass fraction below 0.75, following the proto-GC criterion adopted by \citet{vandonkelaar:2023aa}. All selected systems satisfy this baryon-fraction criterion, as all identified clusters exhibit negligible bound DM content.

Throughout this work, we refer to the identified systems as ``clusters''. However, it is important to note that these objects are not equivalent to the (globular) clusters observed in the local Universe. As discussed later (Section~\ref{sec:results}), most of them are still embedded in and dominated by gas, reflecting an early formation stage rather than a gas-free, dynamically evolved stellar cluster.

\section{Results}\label{sec:results}

As described in \citet{Mayer:2025aa}, the central galaxy in the simulation is a disc-dominated system with a stellar mass of $8 \times 10^{10}$~M$_{\sun}$ and stellar disc radius of about 2~kpc. This galaxy at $z\sim 7.6$ is the result of a major merger of two gas-rich disc galaxies, consistent with vigorously star-forming galaxies above the main sequence at high redshift \citep[see][for a description of its properties and assembly]{mayer:2023aa}. Beyond multiple companion galaxies, we identify multiple stellar clusters within the virial volume of the central main galaxy that are not located inside the galaxies themselves, as indicated in Figure~\ref{fig:showclust1} by the crosses and numbered circles. Thus, whereas in \citet{Mayer:2025aa} the focus was on stellar clusters forming within the discs of these galaxies, here we extend the analysis to clusters emerging along the filaments in the surrounding intergalactic environment.

We identify a total of 55 stellar clusters that satisfy the criteria described in Section~\ref{sec:structures},  all marked in Figure~\ref{fig:showclust1} with crosses and numbered circles. The clusters are located between 2.55~kpc and 23.5~kpc away from the primary halo's centre, the median distance being 6.38~kpc, and have stellar masses between $10^{4.2}$--$10^{7.4}$~M$_{\sun}$. Figure~\ref{fig:showclust2} presents zoomed-in views of the six numbered clusters from Figure~\ref{fig:showclust1}, providing a closer look at their local gas and stellar morphologies.  The maps are centred on the gravitational potential of each system, identified using all particle types in the simulation (gas, stars, and DM\footnote{The bound DM content of all selected clusters is negligible, so the potential is effectively dominated by baryons.}), ensuring a consistent definition of the cluster centre across components. These six clusters were chosen randomly to sample a range of formation environments. In each panel, the left-hand side shows the gas surface density within a (1~kpc)$^2$ region, whereas the right-hand side presents a zoomed-in view of the stellar surface density within a (200~pc)$^2$ region.

\begin{figure}
    \centering
    \includegraphics[ trim={0cm 0cm 0cm 0cm}, clip, width=0.48\textwidth, keepaspectratio]{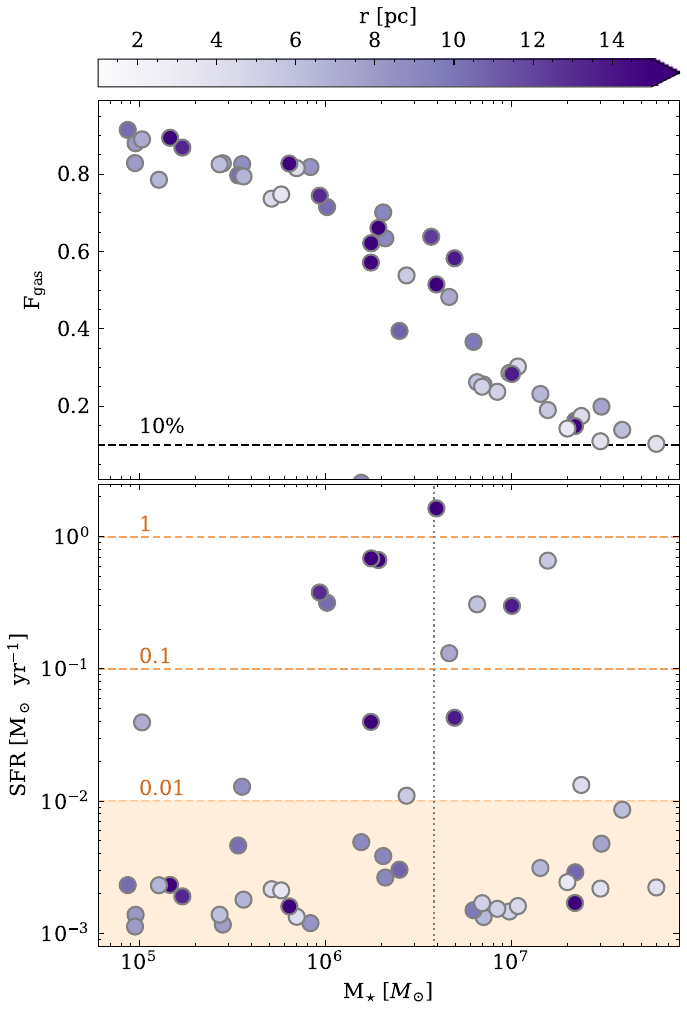}
    \caption{Top panel: gas fraction of the identified stellar clusters as a function of stellar mass. The dashed horizontal line marks a gas fraction of 10 per cent. Bottom panel: average SFR over the past 75~Myr for all identified stellar clusters versus stellar mass. The dashed, orange, horizontal lines indicate SFR levels of 1, 0.1, and 0.01~M$_{\sun}$~yr$^{-1}$, whereas the shaded region highlights clusters with SFRs below 0.01~M$_{\sun}$~yr$^{-1}$. The gray, dotted, vertical line denotes the mean stellar mass of the clusters in the Sunrise Arc at $z \sim 6$ \citep{Vanzell:2023aa}. In both panels, the colour bar represents the distance from the central halo.}
    \label{fig:fgasSFR}
\end{figure}

All clusters in Figure~\ref{fig:showclust2}, except Cluster~6, exhibit a clear peak in gas surface density coinciding with the stellar density peak. The top panel of Figure~\ref{fig:fgasSFR} shows that the vast majority of the 55 analyzed clusters maintain a gas fraction\footnote{Defined as the gas mass divided by the total baryonic mass within twice the half-mass radius of the cluster.} above 10~per cent, indicated by the black dashed line. The gas fraction correlates strongly with cluster stellar mass within the half-mass radius, with the most massive clusters displaying the lowest gas fractions, but shows no apparent dependence on 3D distance from the central halo, suggesting that internal cluster processes, rather than large-scale environmental effects, primarily regulate the gas content.

Again zooming in on the six highlighted clusters, Figure~\ref{fig:showclust2} reveals that the gas distribution may be slightly more centrally concentrated than the stellar component, although this trend is not definitive, consistent with a collapsing component gravitationally bound to the cluster. However, the pronounced gas peaks may be partially influenced by numerical effects inherent to the simulation, such as finite particle resolution or hydrodynamic smoothing, so the precise degree of central concentration should be interpreted with caution. Clusters~2 and 4 are the only clusters in this subsample for which the gas surface density shows a peak only at the cluster location, with no significant gas in the immediate surroundings. In contrast, Clusters~1, 3, 5, and 6 (although Cluster~6 lacks a distinct local gas surface density peak), are embedded within broader gaseous environments that extend beyond the cluster. This illustrates that stellar clusters can be found in a variety of environments, ranging from isolated pockets of gas to dense, extended, gas-rich regions.

In addition, all the identified clusters are strongly baryon-dominated, as expected from our selection criteria, which naturally exclude systems capable of retaining significant DM. The mean baryon fraction within the half-mass radius is 0.92, with a minimum of 0.76. We verified that applying the same mass and distance cuts without the explicit DM requirement does not introduce any DM-dominated systems, confirming that clusters in this regime are intrinsically baryon-dominated rather than filtered by a restrictive DM cut (see Section~\ref{sec:structures}). One can thus conclude, that the forming clusters behave like classical star clusters rather than systems embedded in DM haloes. Their central gravitational potential is set almost entirely by the stellar and gaseous components, supporting the view that they are (proto-)clusters rather than dwarf galaxies or DM-dominated haloes.

To further characterize the evolutionary state of the stellar clusters, we examine their recent SF activity\footnote{The SF rates (SFRs) of the clusters are averaged over 75 Myr to smooth stochastic fluctuations in low-mass clusters. While this time-scale is longer than the instantaneous bursts typically reported in observations, it provides a robust measure of the clusters' overall star-forming activity. Consequently, comparisons to observational SFRs in the upcoming paragraphs should be interpreted with this difference in mind.} over the past 75~Myr. The bottom panel of Figure~\ref{fig:fgasSFR} shows the average SFRs against the stellar mass of the cluster. Most clusters exhibit relatively low SFRs, with the majority of our clusters having SFRs below $\sim$$10^{-2}$~M$_{\sun}$~yr$^{-1}$, whereas a smaller subset remains actively forming stars at higher rates. Again, there is no clear dependence of SFR on distance from the central halo, reinforcing the interpretation that SF is governed primarily by local rather than global environmental conditions.

Next to showing the full distribution, the bottom panel of Figure~\ref{fig:fgasSFR} highlights three characteristic ranges of SFRs. The high-SFR end of the distribution reaches values close to $\sim$1~M$_{\sun}$~yr$^{-1}$, comparable to those inferred for compact star-forming systems such as the clusters detected in the Sunrise Arc at $z \sim 6$, which exhibit global SFRs of order $\sim$1~M$_{\sun}$~yr$^{-1}$ \citep{Vanzell:2023aa}. The vertical, dotted, gray line marks the mean stellar mass of the Sunrise Arc clusters, and the proximity of one cluster in our sample to this value suggests that it occupies a similar mass-SFR regime. An intermediate regime, with SFRs around $0.1$~M$_{\sun}$~yr$^{-1}$, is consistent with values typically observed for young massive clusters forming in dense star-forming environments \citep[see, e.g.][]{Wilson:2006aa, Portegies:2010aa, Elmegreen:2018aa}. Finally, the low-SFR tail of the distribution, below $10^{-2}$~M$_{\sun}$~yr$^{-1}$, corresponds to systems with strongly suppressed ongoing SF. The broad variation indicates that local environmental factors, such as gas density, accretion, and feedback play a key role in regulating the ongoing growth of these systems

\subsection{Cluster properties} \label{sec:cluster_properties}

\begin{figure}
    \centering
    \includegraphics[ trim={0cm 0cm 0cm 0cm}, clip, width=0.48\textwidth, keepaspectratio]{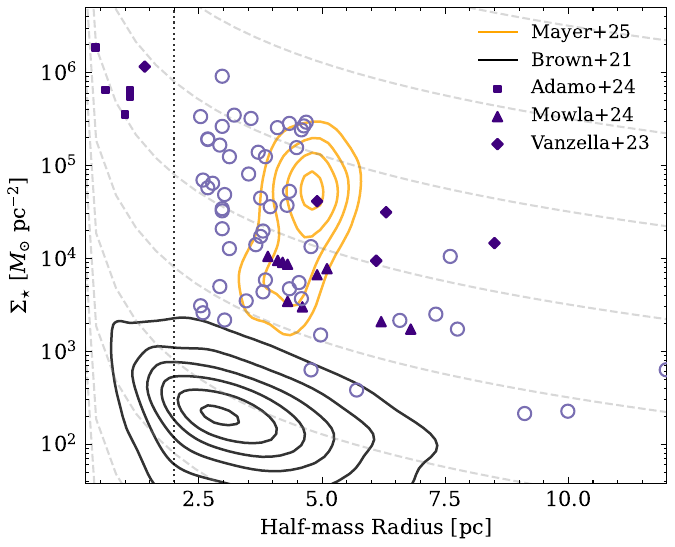}
    \caption{Correlation between stellar surface density, $\Sigma_{\star}$, and half-mass radius of the clusters (open purple circles). Gray dashed lines represent star clusters of equal mass, whereas the black contour lines shows the stellar density-radius relation for clusters in local galaxies from \citet{Brown:2021aa}, using clusters from the LEGUS survey. The dark purple markers represent the stellar surface density and radius of clusters identified in the Cosmic Gems Arc at $z \sim 9.6$ \citep{Adamo:2024aa, Messa:2025aa}, the Firefly Sparkle at $z \sim 8.3$ \citep{Mowla:2024aa}, and the Sunrise Arc at $z \sim 6.0$ \citep{Vanzell:2023aa}. The orange contour lines indicates the clusters found in the galactic discs in the work of \citet{Mayer:2025aa}. Lastly, the vertical, black, dotted line corresponds to the softening length of the simulation.}
    \label{fig:dens}
\end{figure}

We next consider the structural properties of the star clusters by examining the relationship between the stellar surface density and half-mass radius as shown in Figure~\ref{fig:dens}, alongside measurements from the Cosmic Gems Arc at $z \sim 9.6$ \citep{Adamo:2024aa, Messa:2025aa}, the Firefly Sparkle at $z \sim 8.3$ \citep{Mowla:2024aa}, and the Sunrise Arc at $z \sim 6.0$ \citep{Vanzell:2023aa}. The stellar surface densities are calculated at the half-mass radius, consistent with the approach of \citet{Mayer:2025aa}.

Most clusters in our sample exhibit extremely high stellar surface densities, typically 1--3 orders of magnitude greater than those observed in local star clusters found in spiral galaxies, as indicated by the black contour lines \citep{Brown:2021aa}\footnote{We note that in \citet{Mayer:2025aa} this relation was plotted incorrectly. Since there were no conclusions taken from this relation in that work, the main conclusions in \citet{Mayer:2025aa} still stand.}. The densities span a wide range, with several clusters exceeding $10^5$~M$_{\sun}$~pc$^{-2}$, comparable to the dense systems in the Cosmic Gems Arc, whereas others are closer to the values measured in the Sunrise Arc and the Firefly Sparkle.  

The orange contour lines in Figure~\ref{fig:dens} mark the location of the disc clusters identified by \citet{Mayer:2025aa}, which mainly formed through disc fragmentation. The clusters formed within the CGM in our sample show comparable stellar densities and sizes. Nevertheless, we also find clusters that are significantly larger and less dense, as well as smaller and denser, than those identified in the disc by \citet{Mayer:2025aa}, highlighting the diversity of structural properties in our sample.

Still, it is important to note that these density estimates are likely conservative, as the steep central profiles in some clusters suggest values 2--3 orders of magnitude higher. However, the innermost regions of smaller clusters may fall below the gravitational softening length and are therefore not considered further. Additionally, we compare observational effective radii, typically derived from half-light measurements, to the stellar half-mass radii. At these redshifts, stellar clusters are generally quite compact, as shown in Figure~\ref{fig:dens}, meaning that the effective radius is usually smaller than the stellar half-mass radius. Consequently, clusters identified in this work are expected to shift slightly upwards along the gray dashed lines, which represent clusters of equal mass but varying radius.

\subsubsection{Stellar metallicity}\label{sec:metal}

\begin{figure}
    \centering
    \includegraphics[ trim={0cm 0cm 0cm 0cm}, clip, width=0.48\textwidth, keepaspectratio]{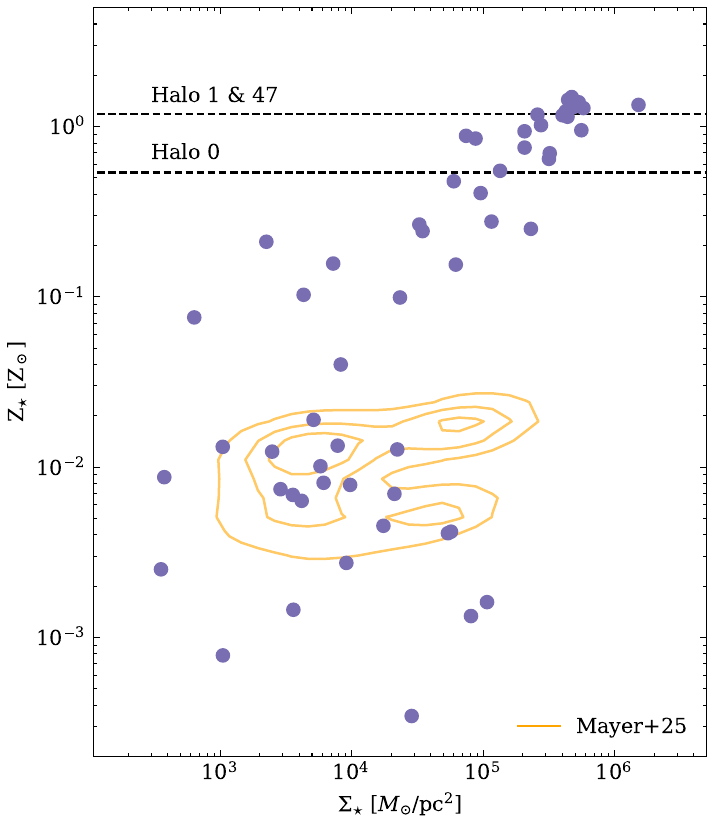}
    \caption{Stellar metallicity as a function of stellar surface density. Purple points mark the clusters identified in this work, while orange contour lines show the distribution of disc clusters from \citet{Mayer:2025aa}.}
    \label{fig:metal}
\end{figure}

Another key property of the cluster population is their stellar metallicity. Figure~\ref{fig:metal} shows the distribution of stellar metallicity, $Z_{\star}$, as a function of stellar surface density for all clusters in our sample. The metallicity of each cluster was determined by computing the mass-weighted average of its member star particles. The clusters span a broad metallicity range, $0.00035 \lesssim Z_{\star}/Z_{\sun} \lesssim 1.49$, with a mean value of $\langle Z_{\star}/Z_{\sun} \rangle \simeq 0.45$. For comparison, we also show the stellar metallicities of the clusters formed within the disc in \citet{Mayer:2025aa}. These disc clusters exhibit a much smaller metallicity spread: $0.0016 \lesssim Z_{\star}/Z_{\sun} \lesssim 0.21$. This indicates  that the disc clusters formed in a more chemically well-mixed environment, while clusters in the outskirts and along filaments sample a more diverse set of physical conditions, leading to a larger spread in stellar metallicities across clusters.\footnote{\textsc{MassiveBlackPS} include subgrid turbulent metal diffusion \citep[][]{Shen:2010aa, Shen:2013aa}, which enables metals to mix between SPH fluid elements, mitigating the well-known tendency of SPH to suppress mixing. This ensures that the observed metallicity spread reflects genuine environmental diversity rather than a numerical artifact.}

Interestingly, the densest clusters in our sample also exhibit the highest stellar metallicities. This correlation likely arises because metal-rich gas cools more efficiently, leading to higher densities and more compact cluster formation. In the simulations, this trend may be further enhanced by the gravitational softening length of 2~pc, which limits how compact the clusters can become and may preferentially affect the lower-metallicity, more diffuse systems. 

We additionally indicate with black, vertical, dashed lines the mean stellar disc metallicities of the main haloes from \citet{Mayer:2025aa}, where Halo~1 and Halo~47 have nearly identical disc metallicities. Notably, the most compact and densest clusters in our sample exhibit metallicities that closely match the characteristic disc metallicities of these systems. One might therefore expect these clusters to be located in the prominent filament connecting Halo~0 with the smaller haloes Halo~1 and Halo~47 (i.e. the thick filament extending upwards in Figure~\ref{fig:showclust1}). However, no such correlation is observed, the distribution of high-metallicity, dense clusters is instead widely scattered and appears largely uncorrelated. At the same time, the high-metallicity clusters are also not confined to being close to the main halo, indicating that enrichment by satellites or pre-enriched inflowing gas also contributes to the observed metallicity distribution.

This apparent randomness reflects the complex interplay of gas accretion, mixing, and local feedback. It shows that high-metallicity gas can be transported and recycled over large distances, and clusters can form from this enriched material in dynamically diverse environments, decoupling their metallicity from the underlying filamentary structure.

\subsection{BH predecessors}\label{bhorgc}

As noted in Section~\ref{sec:cluster_properties}, most clusters in our sample exhibit extremely high stellar surface densities. In such dense environments, it is plausible that IMBHs could form at the centres of the densest clusters through runaway collisions, especially when stellar metallicity is sufficiently low \citep[see, e.g.][]{Portegies:2002aa, Portiegies:2004aa, Portiegies:1999, Devecchi:2009aa, Mapelli:2016aa, Shi:2021aa, Gonzalez:2021aa, Fujii:2024aa, Rantala:2024aa}. For instance, \citet{Fujii:2024aa} showed that densities of $10^3$--$10^4$~M$_{\sun}$~pc$^{-3}$ on parsec scales favour the formation of very massive stars (VMSs). In these regions, stellar mergers dominate over mass loss from winds, producing IMBHs with final masses exceeding $10^3$~M$_{\sun}$. Densities in our clusters are at least comparable to the high-density proto-clusters studied by \citet{Fujii:2024aa}, and in many cases are one to two orders of magnitude higher, suggesting that IMBH formation may be common \citep[see also][]{Rantala:2024aa, Vergara:2025aa}. Moreover, \citet{Fujii:2024aa} find that IMBH mass roughly scales with the square root of the proto-cluster mass. Applying this relation to our most massive cluster (M $\sim 10^{7.4}$~M$_{\sun}$) implies that IMBHs with masses up to $\sim 10^4$~M$_{\sun}$ may form.

The formation of VMSs is most effective at metallicities $Z_{\star} \lesssim 0.05~Z_{\sun}$, where stellar winds are weak \citep[e.g.][]{Sabhahit:2023aa}. VMSs may form from accretion onto a massive star seed or via repeated stellar mergers, provided that mass loss from winds remains low, which again requires low metallicity \citep[e.g.][]{Mapelli:2016aa, Rantala:2024aa}. The exact metallicity range for efficient VMS formation is uncertain, and subsequent IMBH growth could also occur through compact-object mergers or tidal disruptions following the initial seed formation. Within our sample, several clusters could plausibly form IMBHs via this mechanism. While we do not model this formation pathway quantitatively in our simulations, it represents a compelling direction for future studies, wherein our cluster properties and local environments could provide realistic input for $N$-body simulations of runaway collapse \citep[e.g.][]{Katz:2015aa}.

Our sample includes clusters with stellar densities of $10^3$--$10^6$~M$_{\sun}$~pc$^{-3}$, ideal sites for rapid IMBH formation. These IMBHs are expected to form efficiently within these cluster cores and a fraction of these clusters could sink towards the galactic centre, and the IMBHs together with any remaining bound stars from their parent clusters can sink towards the galactic centre via dynamical friction, merging along the way and contributing significantly to the early growth of an SMBH. Given the high density of the background galaxy and the high masses of the clusters, one expects the IMBH-cluster remnants to migrate fast inwards by dynamical friction. Migration will be accompanied by gas inflows, and together they contribute to increase the mass of the central region and assembly of a bulge or dense nucleus \citep[][]{Tamburello_et_al_2017, Dekel:2024aa, vanDonkelaar:2025aa}. The IMBHs that do not contribute to the early growth of the SMBH, or the ones that are dynamically ejected, could be wandering within the galaxy \citep[e.g.][]{Bellovary:2021aa, Ricarte:2021aa, Ricarte:2021ab, Seepaul:2022aa, vanDonkelaar:2025aa}. 

Nevertheless, a large fraction of IMBH-hosting clusters are not expected to sink to the galactic centre, as their dynamical-friction time-scales increase once they lose mass or move into lower-density halo regions (see also Section~\ref{sec:tdf}). Many of these clusters are likely to be partially or fully disrupted during their evolution, leaving behind either a compact stellar remnant or a `naked' wandering IMBH. Some surviving stellar components could resemble the recently identified LRDs. These LRDs are inferred to contain substantial reservoirs of dense gas, which could originate from the original cluster-forming material or be accreted from the surrounding interstellar medium during subsequent evolution in the case of our clusters. Whether such processes can supply enough gas to match the observed properties of LRDs remains uncertain and warrants future investigation. The non-migrating, disrupted IMBH-hosting clusters in our sample could therefore represent plausible progenitors of at least a subset of the LRD population. A detailed study of the properties of these LRD-type clusters, however, lies beyond the scope of the present work.

Additionally, even though there is no guarantee that the clusters that do not form an IMBH would survive to the present day, as some may be tidally disrupted or merge with each other or with the galaxy halo \citep{Forbes:2018aa}, they might be the progenitors of the GCs found in, e.g. the Milky Way. \citet{Gnedin:1997aa} identified relaxation and tidal shocks during disc and bulge passages as the dominant mechanisms driving GC destruction in the MW. Consequently, at small galacto-centric distances, only massive and compact clusters are expected to survive for more than the lookback time (see also Section~\ref{sec:tdf}). Since most clusters in our sample are dense and formed at sufficiently large distances from the central halo, a substantial fraction is likely to survive to the present day.

\subsubsection{Core collapse}

Next to the runaway collision and VMS scenarios mentioned above, core collapse of the cluster could also be a possible pathway to form IMBHs. To explore this pathway further, we estimated the core-collapse time-scales for the cluster population using standard analytical prescriptions for dense stellar systems described in \citeauthor{Dekel:2024aa} (\citeyear{Dekel:2024aa}; see also \citealt{Spitzer:1987aa, Binney:2008aa}). The core-collapse time, $t_{\mathrm{cc}}$ in Myr, was calculated from the two-body relaxation and mass-segregation time-scales as

\begin{equation}
t_{\rm{cc}} = 1.31
\left(\frac{R}{1~\rm{pc}}\right)^{3/2}
\left(\frac{M_{\rm c}}{10^{5}~{\rm M}_{\sun}}\right)^{1/2}
\left(\frac{m_{\rm{mean}}}{10~{\rm M}_{\sun}}\right)^{-1}
\left(\frac{\ln\Lambda}{8.5}\right)^{-1}\,,
\end{equation}

\noindent where $R$ is the cluster half-mass radius, $M_{\rm c}$ is the total baryonic cluster mass, and $\ln\Lambda$ is the Coulomb logarithm. The half-mass radius $R$ is measured directly from the stellar distribution of each cluster, and we adopt a mean stellar mass of $m_{\rm mean} = 0.5~{\rm M}{\sun}$, consistent with a \citet{Kroupa:2001aa} initial mass function (IMF). To account for internal rotation, we apply a reduction factor $f{\rm rot}$ that depends on the ratio of ordered to random motions, such that $t_{\rm cc,rot} = t_{\rm cc} / f_{\rm rot}$, where the rotation parameter is defined as $F_{\rm rot} = V_{\rm rot}^2 / V_{\rm c}^2$, the ratio of rotational to gravitational support. The corresponding seed BH mass expected from rapid core collapse can then be estimated, as described in \citet{Dekel:2024aa}, as

\begin{equation}
M_{\rm{BH}} = 6.8\times10^{-3}\,M_{\rm c}\,\left(\frac{\ln\Lambda}{8.5}\right)\,.
\end{equation}

\begin{figure}
    \centering
    \includegraphics[ trim={0cm 0cm 0cm 0cm}, clip, width=0.48\textwidth, keepaspectratio]{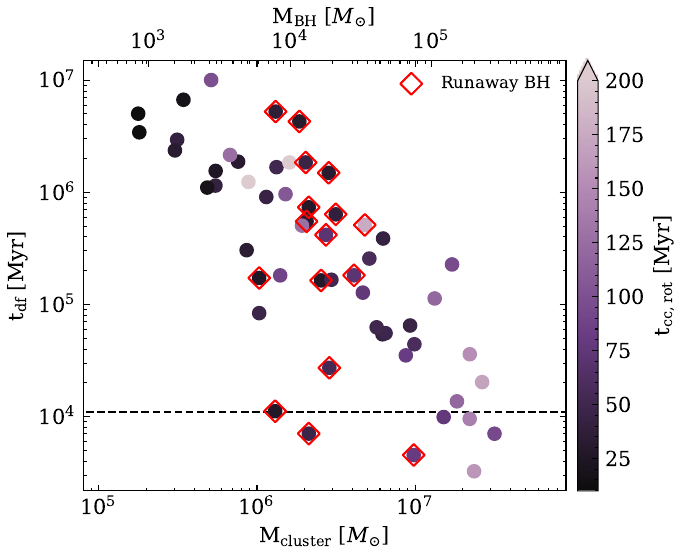}
    \caption{The dynamical-friction time-scale plotted against the total baryonic mass of the clusters. The corresponding seed BH mass assuming the core-collapse formation pathway can be found on the upper $x$-axis. The colour bar represents the expected core-collapse time-scale calculated using the formalism described in \citet{Dekel:2024aa}. Markers encircled by the red diamond are clusters with a stellar density above $10^4$~M$_{\sun}$ pc$^{-2}$ and a $Z_{\star} < 0.5~Z_{\sun}$. The horizontal, black, dashed line shows the lookback time at $z \sim 7.6$.}
    \label{fig:Mtcc}
\end{figure}

Figure~\ref{fig:Mtcc} shows the total baryonics cluster mass along the lower $x$-axis and the corresponding expected seed BH mass along the upper $x$-axis. The colour of each marker indicates the core-collapse time-scale corrected for rotation, $t_{\rm cc,rot}$. Notably, $t_{\rm cc,rot}$ does not correlate clearly with total cluster mass, rather it appears primarily determined by the cluster’s internal structure, particularly its central stellar density. Clusters with core-collapse time-scales much larger than 1~Myr are expected to experience significant supernova (SN) feedback before core collapse, which can disrupt the collapse and affect SF if the gas reservoir has not yet been fully converted into stars. In our sample, the shortest core-collpase time-scale is approximately 10~Myr, suggesting that all clusters are susceptible to SN feedback prior to collapse. We note, however, that these relatively long core-collapse time-scale values are likely influenced by numerical softening, which prevents clusters from reaching smaller, more realistic sizes.

Consequently, the runaway-collision pathway described above appears to be the more plausible formation channel for our BH seeds. This is further illustrated in Figure~\ref{fig:Mtcc}, where clusters with central stellar surface densities exceeding $10^4$~M$_{\sun}$~pc$^{-2}$ and stellar metallicities below $Z_{\star} < 0.5~Z_{\sun}$\footnote{We adopt $Z_{\star} = 0.5Z_{\sun}$ as the upper metallicity limit for descriptive purposes, even though most studies find runaway-collision IMBH formation efficient only at $Z_{\star} \lesssim  0.2Z_{\sun}$ \citep[e.g.][]{Mapelli:2016aa, Rantala:2024aa}. This choice allows us to include all clusters for completeness and to account for possible outliers in cluster properties or extreme environments that could, in principle, host IMBH formation at somewhat higher metallicities than typically assumed.}, i.e. those more likely to form IMBHs through runaway collisions, are highlighted with red diamonds.

\subsubsection{Dynamical friction}\label{sec:tdf}

Figure~\ref{fig:Mtcc} also provides an approximate estimate of the dynamical friction time-scale, i.e. the expected duration for the clusters to spiral into the centre of the main halo of the simulation. The dynamical-friction time-scale, $t_{\rm{df}}$, was derived using the combined density profile of gas, stars, and DM to obtain the enclosed mass $M(<r)$ of the galaxy and circular velocity $V_{\rm c} = \sqrt{G\,M(<r)/r}$ following the formulation of \citet{Binney:2008aa}:

\begin{equation}
t_{\rm{df}} = 1.17\,\frac{V_{\rm c}\,r^2}{G\,M_{\rm c}\,\ln\Lambda}\,.
\end{equation}

\noindent  where $G$ is the gravitational constant. This provides a rough estimate of the time required for clusters on approximately circular orbits to migrate towards the galactic centre due to dynamical friction against the background potential. In reality, radial motions could reduce $t_{\rm df}$, allowing clusters to reach the centre more quickly. From Figure~\ref{fig:Mtcc}, we find that most clusters are unlikely to reach the galactic centre of the main halo within a lookback time (indicated by the black dashed horizontal line), although this conclusion is less certain for clusters on more radial trajectories. Importantly, several massive dwarf galaxies are also located within the main halo’s virial radius. For example, Halo~1, with a total mass of $\sim$$10^{10}$~M$_{\sun}$ \citep[described in][]{Mayer:2025aa}, has a dynamical friction time-scale of $t_{\rm df} \approx 4 \times 10^3$~Myr. Clusters that happen to lie along Halo~1’s trajectory could be swept along and carried into the main halo together with it.

Using the current estimates of $t_{\rm df}$ and the seed BH masses derived from the core-collapse method, we infer a total contribution of $\sim$$10^{4.5}$~M$_{\sun}$ to the central BH of the main halo. If instead we consider 20 per cent of the total stellar mass of the clusters, which provides a more appropriate proxy for the BH mass formed through the runaway-collision scenario \citep[e.g.][]{Fujii:2024aa}, the total potential contribution could reach $\sim$$10^{6.5}$~M$_{\sun}$. As a result, the SMBH would not reach the extreme masses observed in bright, high-redshift quasars without substantial additional growth via gas accretion. Nevertheless, its mass would be comparable to the SMBHs recently discovered by JWST \citep[e.g.][]{Harikane_et_al_2023, Larson_et_al_2023, Maiolino_et_al_2023} and similar to predictions from super-Eddington accretion models acting on light BH seeds in proto-galaxies \citep[e.g.][]{Sassano_et_al_2023, Zana_et_al_2025}. In addition, the potential for high merging activity of IMBHs in dense clusters makes these systems promising sources of gravitational-wave events detectable by the Laser Interferometer Space Antenna \citep[LISA;][]{Amaro-Seoane_et_al_2023, Colpi_et_al_2024}. However, these estimates should be interpreted with caution, as small-scale hardening processes, three-body interactions, and gravitational-wave recoil could significantly delay mergers or even eject BHs from the system \citep[e.g.][]{Holley-Bockelmann:2008aa, Mayer:2013aa}.

\subsection{Filament fragmentation}

The next step is to investigate the formation of the clusters, which we hypothesize arises from gravitational fragmentation of the filamentary structures between the galaxies. We note that this is slightly different from the mechanism of cold stream fragmentation proposed by \citet{Mandelker:2018aa}, which is purely hydrodynamical, being driven by the non-linear growth of the Kelvin-Helmholtz instability along cold supersonic streams.  To assess whether the gas at the cluster locations was gravitationally unstable, we compute the critical mass for collapse, defined as

\begin{equation}
M_\mathrm{crit} = \frac{2 c_{\rm s}^2}{G}\,,
\end{equation}

\noindent where $c_{\rm s}$ is the local thermal sound speed. The criterion provides an estimate of the minimum mass required for a region to collapse under its own gravity, following the methodology used by \citet{Fu:2025aa} to study the formation of free-floating planetary-mass objects along tidally induced filaments forming between interacting  circumstellar discs. It is important to note, however, that this is a simple, local estimate of the minimum mass for collapse. Unlike the classical \citet{Jeans1902} mass, it does not explicitly depend on density and assumes an idealized geometry. A more rigorous analysis in future studies would need to account for density gradients, filament width, and external pressure.

To quantify the instability of the gas that eventually formed each cluster, we calculate the ratio $M_\mathrm{gas}/M_\mathrm{crit}$, where $M_\mathrm{gas}$ is the mass of gas present  at the cluster location 1~Myr prior to the final snapshot. The gas mass is measured within a sphere of radius 100~pc centred on the final cluster position.  While the mean free-fall time of the gas within these regions is $\sim$50~Myr, the densest substructures are expected to have much shorter free-fall times ($\sim$1--3~Myr). Thus, a 1~Myr interval captures the state of the gas immediately preceding collapse while minimizing contamination from lower-density material.

\begin{figure}
    \centering
    \includegraphics[ trim={0cm 0cm 0cm 0cm}, clip, width=0.48\textwidth, keepaspectratio]{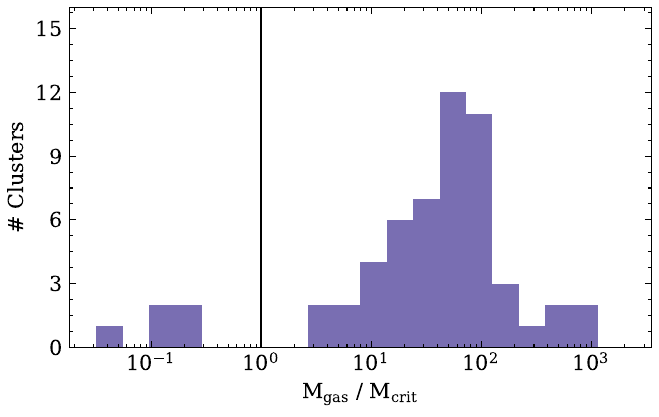}
    \caption{Distribution of $M_\mathrm{gas}/M_\mathrm{crit}$ for all identified stellar clusters. The vertical black line marks $M_\mathrm{gas}/M_\mathrm{crit} = 1$, which represents the threshold for gravitational instability.}
    \label{fig:Mcrit}
\end{figure}

While the critical mass is  calculated using the local thermal sound speed,  which accounts only for thermal support, the gas may also have turbulent or bulk motions. Nevertheless, the typical velocity dispersion in these regions is small compared to the sound speed, so the contribution to support against gravity is minor and does not significantly affect our results. Typical displacements due to bulk motion over 1~Myr are much smaller than the selected radius of a 100 pc, so the fixed-sphere approximation provides a robust estimate of the local gas reservoir. As shown in Figure~\ref{fig:Mcrit}, most clusters have $M_\mathrm{gas}/M_\mathrm{crit} > 1$, indicating that the gas available at that time was sufficiently massive to undergo gravitational collapse.

\begin{figure}
    \centering
    \includegraphics[ trim={0cm 0cm 0cm 0cm}, clip, width=0.48\textwidth, keepaspectratio]{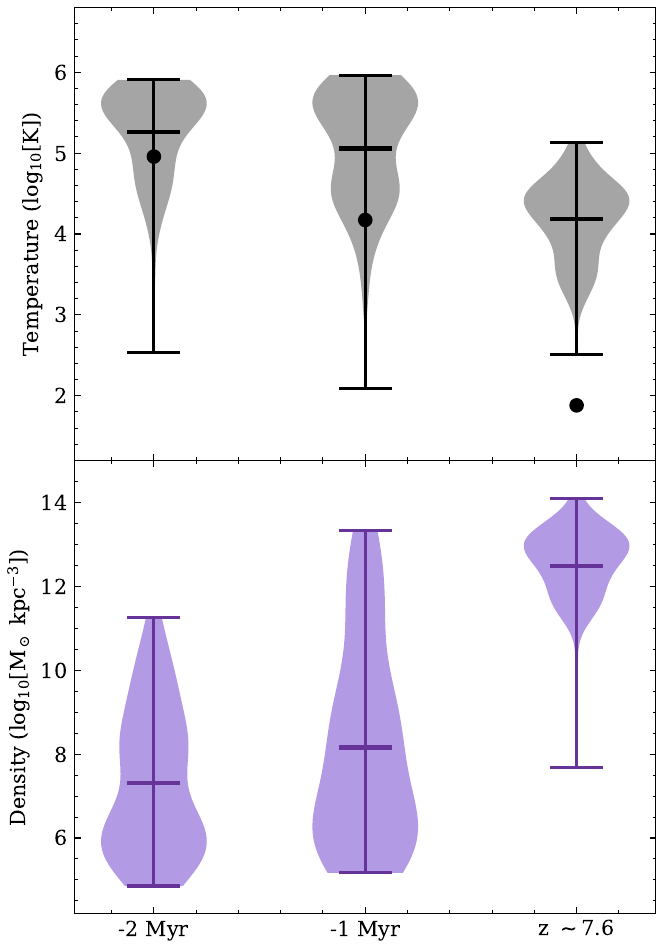}
    \caption{Distributions of gas temperature (top panel) and density (bottom panel) within 100-pc spheres centred on the clusters at three cosmic times: 2~Myr before, 1~Myr before, and at the final snapshot ($z \sim 7.6$), shown from left to right. Gray and purple violins represent the logarithmic temperature and density distributions, respectively. Thick lines inside the violins indicate the mean values, whereas the black dots in the top panel mark the mean of the minimum temperatures across the clusters.}
    \label{fig:frag}
\end{figure}

The low values of $M_\mathrm{gas}/M_\mathrm{crit}$, in particular those of the five clusters with ratios below unity, are predominantly associated with lower stellar mass systems. These clusters have stellar masses in the range $10^{5.5}$--$10^{6.3}$~M$_{\sun}$ and exhibit high gas fractions, as highlighted in Figure~\ref{fig:fgasSFR}. Amongst these five, we also find two of the clusters examined in detail in Figure~\ref{fig:showclust2}, namely Clusters~2 and~3. This indicates that cluster formation in our sample is not limited to regions that satisfy this simple collapse criterion. More generally, collapse in filamentary gas can depend on additional conditions beyond spherical instability \citep[e.g. the line mass, see][]{Mandelker:2018aa}, reinforcing that $M_\mathrm{gas}/M_\mathrm{crit}$ provides only a simplified description of the local collapse conditions.

To further investigate the physical conditions of the gas that gave rise to the clusters, Figure~\ref{fig:frag} presents the distributions of temperature (top panel) and volume density (bottom panel) measured within a sphere of radius 100~pc centred on the clusters' locations in the final snapshot, consistent with the method used to calculate  $M_\mathrm{gas}/M_\mathrm{crit}$. The distributions are shown at three cosmic times: 2~Myr before, 1~Myr before, and at the final snapshot (around $z \sim 7.6$), arranged from left to right. In both panels, the thick lines inside the violins indicate the mean values of the distributions. In the top panel, the dots show the average minimum temperature across all clusters\footnote{Computed as $\langle T_\mathrm{min} \rangle = \frac{1}{n} \sum_{i=1}^{n} T_{\rm min, i}$\,.}.

We find that, over time,  the gas in these regions cools and contracts, with the temperature decreasing, particularly the mean minimum temperature (the black dots)and the density increasing. These trends reinforce the conclusion that the gas in the vicinity of the clusters was sufficiently dense and cool to undergo gravitational collapse, consistent with the filament fragmentation scenario and the $M_\mathrm{gas}/M_\mathrm{crit} > 1$ analysis.

\subsubsection{Gas metallicity}

\begin{figure}
    \centering
    \includegraphics[ trim={0cm 0cm 0cm 0cm}, clip, width=0.48\textwidth, keepaspectratio]{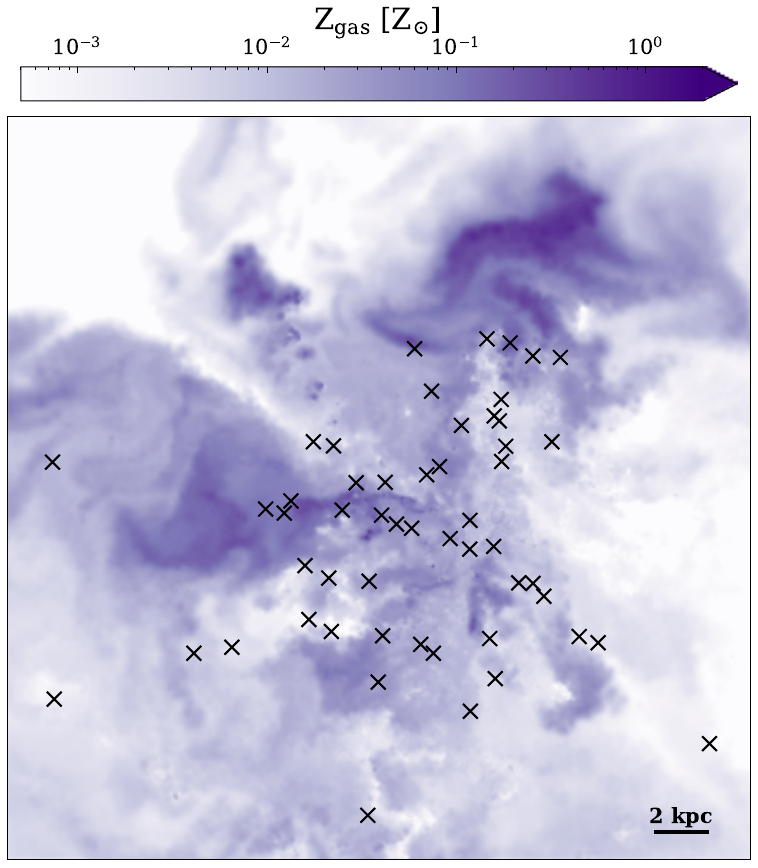}
    \caption{Metallicity distribution in the gas streams and central halo of the simulated system at the final snapshot, $z \sim 7.6$. The colour scale shows the mass‑weighted mean gas metallicity $Z_{\rm gas}$ projected along the line-of-sight. The black crosses mark all identified clusters, similarly to Figure~\ref{fig:showclust1}.}
    \label{fig:Zgas}
\end{figure}

To better characterize the chemical environment in which the clusters formed, Figure~\ref{fig:Zgas} shows the spatial distribution of gas metallicity in the final snapshot at $z \sim 7.6$. It seems that clusters located farther from the main halo tend to reside in regions of low to intermediate metallicity, although this trend is weak, as many clusters are also found in more metal-rich areas. Overall, these metallicity levels indicate only moderate enrichment of the surrounding gas, likely dominated by recently accreted intergalactic material rather than recycled outflows. Such conditions are broadly consistent with cosmological simulations of cold-stream fragmentation \citep[e.g.][]{Mandelker:2018aa}, in which star-forming clumps arise within metal-poor filaments feeding massive high-redshift galaxies. The range in $Z_{\rm gas}$ suggests that enrichment and SF occurred nearly simultaneously, before feedback or turbulent mixing could potentially homogenize the metallicity of the surrounding medium.

\begin{figure}
    \centering
    \includegraphics[ trim={0cm 0cm 0cm 0cm}, clip, width=0.48\textwidth, keepaspectratio]{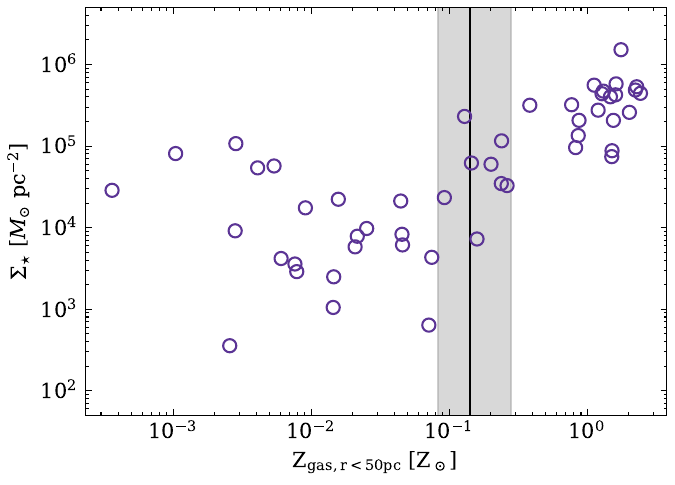}
    \caption{Stellar surface density, $\Sigma_{\star}$, as a function of the mean gas metallicity, $Z_{\rm gas}$, within a 50-pc sphere around each cluster. The black vertical line indicates the mean gas metallicity of the simulation (excluding the main halo) with the standard deviation shown in gray. }
    \label{fig:Zdens}
\end{figure}

Figure~\ref{fig:Zdens} shows the stellar surface density, $\Sigma_{\star}$, as a function of the mean gas metallicity, $Z_{\rm gas}$, measured within a 50-pc sphere around each cluster, which captures the proto-cluster environment while minimizing contamination from the surrounding gas. The black vertical line indicates the mean gas metallicity outside the main halo and serves as a reference for the typical background environment. We find that the densest clusters form in regions with metallicities higher than the global average, as expected from Figure~\ref{fig:metal} and Section~\ref{sec:metal}, suggesting that an enriched environment would promote the formation of compact stellar systems. Higher initial metallicity can promote more compact gas and efficient SF, while overdense collapsing regions may also form stars rapidly, enriching the surrounding gas and accelerating further collapse. Additionally, enhanced metallicity increases cooling, allowing gas to reach higher densities before feedback disperses it, which favours the formation of compact, bound clusters \citep[e.g.][]{Fukshima:2020aa}. In contrast, in galactic centres strong merger-driven compression can achieve comparable densities even at low metallicity, providing an alternative route to dense cluster formation, as seen in \citet{Mayer:2025aa}, where very dense clusters arise from disc fragmentation (see also Figure~\ref{fig:metal}).

Lower-density clusters are generally associated with lower-metallicity gas, indicating formation in less chemically evolved regions that have experienced fewer enrichment and feedback cycles, though the stellar metallicities of the clusters themselves were not compared.  Reduced cooling efficiency in such environments likely leads to more extended star-forming regions and lower stellar densities. This metallicity dependence provides insight into the environmental conditions regulating cluster formation and helps explain the diversity of cluster structural properties across cosmic time.

\subsubsection{BH predecessor environment}\label{sec:form2}

\begin{figure}
    \centering
    \includegraphics[ trim={0cm 0cm 0cm 0cm}, clip, width=0.49\textwidth, keepaspectratio]{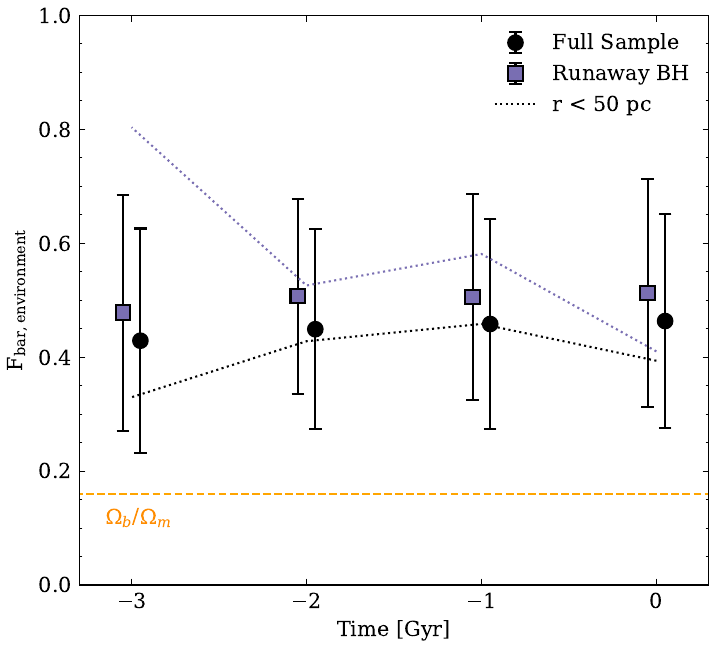}
    \caption{Evolution of the baryon fraction in the environments where different types of clusters form. Purple points mark runaway BH clusters, defined as systems with stellar densities above $10^4$~M$_{\sun}$ pc$^{-2}$ and stellar metallicities below 0.5~$Z_{\sun}$, whereas black points show the full cluster sample for comparison. Markers indicate the baryon fraction measured within a 0.5~kpc radius sphere, dotted lines show the evolution within a 50~pc radius. The error bars represent the standard deviation of the baryon fraction at each time step, calculated across the selected clusters. On the $x$-axis, `Time = 0' corresponds to $z \sim 7.6$. Lastly, the orange dashed line shows the cosmic baryon fraction.}
    \label{fig:fbar}
\end{figure}

To provide broader context and examine the differences between clusters that are likely to form runaway BHs and those that are not, Figure~\ref{fig:fbar} presents the evolution of the baryon fraction, $F_{\rm bar}$, in their respective environments. It compares clusters associated with runaway-collision BHs to the full sample of clusters, as shown in Figure~\ref{fig:Mtcc}. The baryon fraction is computed as

\begin{equation}
F_{\rm bar} = \frac{M_{\star} + M_{\rm gas}}{M_{\star} + M_{\rm gas} + M_{\rm DM}}\,.
\end{equation}

The mass is calculated within a sphere centred on the centre of mass of the stars that will eventually form the clusters in the final snapshot. Solid markers indicate values within a radius of 0.5~kpc, while dotted lines show $F_{\rm bar}$ measured within a smaller radius of 50~pc. We find that the baryon fraction within 0.5~kpc exhibits little evolution over time for either population, indicating that the large-scale environments of both cluster types are relatively stable. This suggests that the kpc-scale baryon fraction alone is not a strong discriminator between runaway-BH progenitors and surviving proto-GCs.  In both cases, the fraction lies above the cosmic average of $\sim$0.16, which is expected for gas-rich filamentary environments, where baryons are naturally concentrated by large-scale inflows. Within a larger 2.5~kpc diametre, the values decrease to $\sim$0.17, suggesting that the higher values at smaller radii simply reflect local baryon concentration within the filaments.

The difference might be small, but the clusters that have the properties to facilitate formation of runaway BHs (as shown in Figure~\ref{fig:Mtcc}) systematically occupy regions with slightly higher $F_{\rm bar}$ compared to the full sample, indicating that they tend to form in less DM-dominated regions of the filaments. Importantly, this offset is small on the bigger scale, and only becomes significant on scales of $\lesssim 50$~pc, where the baryon fraction shows much stronger temporal variability. This scale dependence suggests that the decisive physical differences arise from small-scale, stochastic fluctuations in gas density and temperature within the filaments, which can be non-linearly amplified during collapse. 

Focussing further on the small scale evolution (within 50~pc), at $\sim$3~Myr before the final snapshot, the clusters that have the potential to form runaway BHs lie in clearly more baryon-dominated environments, with $F_{\rm bar}$ well above that of the full sample. Shortly afterwards, their $F_{\rm bar}$ decreases rapidly, reflecting prompt local baryonic collapse and the ensuing burst of SF. Then, as the gas is converted into stars and some of the loosely bound material is removed through dynamical interactions, tidal stripping, or stellar feedback (e.g. SNae and stellar winds), the initially enhanced baryon fraction declines (see Section~\ref{sec:method} for the baryonic physics implemented in the simulation). By the time the clusters reach close to their peak stellar mass, both populations exhibit similar $F_{\rm bar}$ values within 50~pc, suggesting that local baryonic collapse within the filaments ultimately erases most of the initial environmental differences. 

As a result, one could conclude that proto-GCs, i.e. the potentially surviving star clusters without a runaway BH, tend to form in regions that remain more DM-influenced on these small scales, where extreme baryonic runaway is avoided. In this context, the formation of these clusters bears some similarity to the SIGO scenario described by \citet{Lake:2022aa}, in which converging gas flows along filaments generate local gravitational instabilities near DM overdensities, enabling baryonic collapse outside the main bodies of DM haloes. In our case, however, the clusters form within a filament connecting galaxies rather than in the streaming-velocity-dominated environments characteristic of SIGOs. The broad notion of collapse occurring outside haloes but still influenced by nearby DM structure is shared. These results also align with our previous findings in \citet{vandonkelaar:2023aa}, wherein proto-GCs form within filamentary structures and are located within $\sim$5~kpc of nearby DM subhaloes, highlighting the key role of filamentary inflows and the surrounding DM environment in shaping early cluster formation.

This distinct $F_{\rm bar}$ evolution suggests that the runaway outcome is not primarily set by large-scale baryon dominance, but is highly sensitive to small-scale perturbations on tens-of-parsec scales. The small pre-collapse differences in density, temperature, or metallicity within the filaments can therefore turn an otherwise similar region into the runaway collapse regime, naturally leading to a degree of stochasticity in IMBH progenitor formation. In particular, enhanced metallicity can strongly increase radiative cooling efficiency, making metal-rich regions within the filaments more susceptible to thermal instability prior to the onset of full gravitational collapse. Such thermally driven collapse amplifies small density and temperature fluctuations, seeding the non-linear runaway that produces extreme central densities. In such cases, the resulting high densities favour runaway stellar collisions, ultimately leading to the formation of a massive BH that consumes or disrupts much of the surrounding stellar component.  Consequently, these systems are unlikely to survive as long-lived clusters, persisting mainly through their BH remnant. By contrast, clusters that avoid this runaway remain in a regime of more moderate baryonic assembly and may survive as bound (proto-)GCs.

\section{Discussion and conclusions}\label{sec:disc}

The formation of compact stellar systems outside galactic discs at $z > 7$ provides an alternative route to early cluster assembly. In this work, we have shown that dense star clusters can arise directly within the filamentary structures of the CGM, forming efficiently through local gravitational instabilities in gas that has cooled and condensed during filament accretion. These systems, the ``cosmic wallflowers'', exhibit stellar surface densities up to $\Sigma_{\star} \gtrsim 10^5$ M$_{\sun}$~pc$^{-2}$ and stellar masses of $10^{4.2}$--$10^{7.4}$~M$_{\sun}$, values that are fully consistent with the compact clusters recently observed by JWST in strongly lensed galaxies such as the Cosmic Gems Arc, the Sunrise Arc, and the Firefly Sparkle \citep{Vanzell:2023aa, Mowla:2024aa, Adamo:2024aa}.

The filament fragmentation mechanism identified here differs from the classical picture of cluster formation in turbulent galactic discs. In the environment of \textsc{MassiveBlackPS}, the converging flows within dense filaments lead to rapid collapse across multiple regions, giving rise to numerous isolated clusters within a few Myr. Most clusters satisfy $M_{\rm gas}/M_{\rm crit} > 1$, confirming that their birth sites were gravitationally unstable prior to collapse. The accompanying decline in temperature and rise in density further support the scenario in which thermal instability and gravitational compression together drive fragmentation of the inflowing filaments. These findings have similarities with the predictions of the models presented by \citet{Lake:2022aa}, as well as with the proto-GC results reported in \citet{vandonkelaar:2023aa}.

Additionally, we find that gas metallicity plays an important role in modulating these processes. The densest clusters form in regions with metallicities above the global mean, where enhanced cooling allows the gas to reach very high densities before SN feedback becomes effective, though this result is sensitive to the specific implementation of stellar feedback in the simulation. This promotes the formation of compact bound systems and increases the likelihood of IMBH formation through runaway stellar collisions. In contrast, metal-poor gas cools less efficiently, leading to clusters with lower central densities that never enter the runaway regime and are therefore more likely to survive as long-lived (proto-)GCs. These metallicity-driven differences, combined with local variations in gas density and DM content, naturally give rise to the two evolutionary pathways identified in Section~\ref{sec:form2}: collapse into an IMBH remnant versus survival as a stellar cluster. These pathways might not be mutually exclusive: some clusters may occupy an intermediate regime, as suggested by systems like Omega Centauri, whose cluster nature as well as the central BH is still debated \citep[][]{Baumgardt:2019aa, Chen:2025aa, Banares:2025aa}.

We note that the models used to estimate IMBH formation via runaway collisions \citep[e.g.][]{Fujii:2024aa, Rantala:2024aa} do not account for the probability of IMBH ejection from their host clusters. Depending on the cluster dynamics, some IMBHs may be ejected at velocities that allow them to wander in the galactic halo or escape entirely. Consequently, the actual contribution of these IMBHs to the central SMBH mass could be lower than the upper-limit estimates presented here. Applying the scaling from \citet{Fujii:2024aa} yields central IMBH masses up to $\sim$$10^4$~M$_{\sun}$ for the most massive clusters, making these systems potential building blocks of early SMBHs. Dynamical friction will cause the most massive clusters to migrate towards the galactic nucleus within a Hubble time, where their IMBHs could merge or accrete, contributing to the early growth of central SMBHs \citep[e.g.][]{Tamburello_et_al_2017,Dekel:2024aa}. The less massive and farther away systems, by contrast, may remain wandering IMBH hosts in the halo, connecting the cluster population explored here with the population of off-centre BHs \citep[e.g.][]{Lin:2018aa, Yao:2025aa, vanDonkelaar:2025aa}. The remaining clusters, which are either less dense or lie at the higher end of the metallicity range, have a strong likelihood of surviving to the present day and may represent proto-GCs, although this interpretation remains tentative. While higher metallicities generally correlate with higher pre-collapse gas densities in our simulation, very high metallicities also imply stronger stellar winds and mass loss \citep[see, e.g.][]{Mapelli:2016aa, Rantala:2024aa}, which can suppress IMBH formation even in otherwise compact clusters. Given that most clusters in this sample are dense and formed far from the central halo, a substantial fraction is nevertheless expected to persist to the present.

Thus, filament fragmentation could provide a natural explanation for the offset clusters seen by JWST \citep{Mowla:2022aa, Claeyssens:2023aa, Giunchi:2025aa, Ong:2025aa} and for the presence of massive BHs at high redshift. The small nuances in the birth environments of the clusters highlight the role of local conditions in shaping their evolution. The IMBH progenitors produced through runaway stellar collisions form in regions with weaker DM dominance, allowing the gas to reach extremely high densities and low temperatures on short time-scales. Under these conditions, the stellar system collapses rapidly and most of its baryonic mass is consumed or disrupted during the runaway phase, so these objects do not survive as long-lived clusters, but instead leave behind a massive BH remnant. Proto-GCs, by contrast, form in more DM-rich environments with lower baryon fractions, where the slower collapse and reduced cooling efficiency prevent them from entering the runaway regime and allow the stellar cluster to survive. Metallicity differences further emphasize this separation, with the densest clusters exhibiting higher metallicities that enhance cooling and accelerate the path towards runaway collapse.

Nevertheless, it is important to stress that filaments might be warmer and less prone to fragmentation if the simulation had included local UV radiation from star-forming regions. The clusters discussed form at distances of $\sim$2.5~kpc to $\sim$23.5~kpc (median $\sim$6.4~kpc) from the central galaxy, whose recent SFR is about $147.5 $~M$_{\sun}$~yr$^{-1}$. A simple geometric estimate following a similar same approach as \citet{Baumschlager:2025aa}, shows that if ionising photons escaped freely, the resulting local UV flux at these distances would be higher for the clusters than the mean cosmic UV background employed \citep[][]{Haardt:2012aa}. In reality, only a fraction of the ionising radiation can reach the filaments, because of absorption and scattering within the interstellar medium and CGM. The radiative-transfer study by \citet{Baumschlager:2025aa} demonstrates that, when radiative-transfer effects are treated self-consistently, local radiation does not heat the gas evenly. This allows cluster-forming filaments to remain cold. We note, however, that our conclusions are also sensitive to the treatment of small-scale SF and feedback, which can influence cluster densities, survival, and the availability of gas in the filaments. Overall, the general picture that the cluster-forming filaments cool and condense should therefore be still consistent with expectations, once proper radiative-transfer effects are taken into account. Nonetheless, this represents an important direction for future studies.

In conclusion, we find that filament fragmentation is a viable and efficient mechanism for producing dense stellar clusters in the early Universe, generating star clusters consistent with the lensed clusters observed at high redshift. The local metallicity of the gas is also a key factor shaping the density, compactness, and evolutionary pathways of these clusters, with metal-rich regions tending to favour the formation of denser systems. This formation process provides an alternative to in-disc fragmentation and is likely most effective in highly overdense regions, where strong accretion flows drive rapid cluster formation. This would imply that a significant fraction of today's GCs, and possibly some of the early seeds of the SMBHs, originated as ``cosmic wallflowers'': compact, gravitationally bound clusters born in the quiet outskirts of forming galaxies rather than within their bright discs. Upcoming JWST and ALMA observations targeting the extended environments of high-redshift galaxies will be able to provide critical tests of this picture by searching for compact, off-disc clusters embedded in cold filamentary gas.

\section*{Acknowledgements}
FvD acknowledges support from the Herchel Smith Fellowship at the University of Cambridge. LM acknowledges support from the Swiss National Science Foundation under the Grant 200020\_207406. PRC and FvD acknowledge support from the Swiss National Science Foundation under the Sinergia Grant CRSII5\_213497 (GW-Learn). DS acknowledges support from the Science and Technology Facilities Council (STFC) under grant ST/W000997/1. AA acknowledges support by the Swedish research council Vetenskapsr{\aa}det (VR) project 2021-05559, and VR consolidator grant 2024-02061.

\section*{Data Availability}
The data underlying this article will be shared on reasonable request to the corresponding author.



\bibliographystyle{mnras}
\bibliography{example} 


\bsp	
\label{lastpage}
\end{document}